\documentclass[aps,showkeys,nofootinbib]{revtex4}
\usepackage[english]{babel}
\usepackage[utf8]{inputenc}
\usepackage[T1]{fontenc}
\usepackage{amsmath}
\usepackage{amsfonts}
\usepackage{amssymb}
\usepackage{amsthm}
\usepackage{graphicx}
\usepackage{array}
\usepackage{dcolumn}
\usepackage{epstopdf}
\usepackage{multirow}
\usepackage{bm}
\usepackage{braket}
\usepackage{xcolor}

\begin{document}

\title{Perturbative results of fractional quantum mechanics for two particular systems}

\author{Claude \surname{Semay}}
\email[E-mail: ]{claude.semay@umons.ac.be}
\thanks{ORCiD: 0000-0001-6841-9850}

\author{Clara \surname{Tourbez}}
\email[E-mail: ]{clara.tourbez@umons.ac.be}
\thanks{ORCiD: 0009-0004-0909-6974}

\author{Lo\"ic \surname{Keszeli}}
\email[E-mail: ]{Loic.Keszeli@student.umons.ac.be}

\affiliation{Service de Physique Nucl\'{e}aire et Subnucl\'{e}aire,
Universit\'{e} de Mons,
UMONS Research Institute for Complex Systems,
Place du Parc 20, 7000 Mons, Belgium}

\date{May 2026}

\begin{abstract}
The fractional Schr\"odinger equation is studied with a kinetic energy that slightly deviates from the usual nonrelativistic form. The harmonic oscillator and the Kepler problem are both treated in the context of small perturbations. The usual perturbation theory is used and compared with the envelope theory. The corrections to the unperturbed Hamiltonians are analytical. The bounds computed with the envelope theory are good for the harmonic oscillator but not quite as good for the Kepler problem. \end{abstract}


\maketitle

\section{Introduction}
\label{sec:intro}

Microscopic physics is very well described by the ``usual'' quantum mechanics in $3D$ space relying on the time-dependent Schr\"odinger equation which is of first order in time derivative and of second order in space derivatives. Fractional quantum theories extend standard quantum mechanics by replacing either the spatial operator, the time derivative, or even the dimensionality of the underlying space with fractional generalizations.

Time-fractional evolution modifies the Schr\"odinger equation by replacing the first-order time derivative with a fractional derivative, usually of Caputo type \cite{nabe04,torr24}. The wave function then possesses memory effects, since in this case the evolution depends on the system's past history. Solutions are typically expressed through Mittag-Leffler functions rather than simple exponentials, and the dynamics may become non-Markovian, making this framework useful for complex and dissipative quantum environments.

Space-fractional quantum mechanics, introduced by Laskin, replaces the Laplacian in the Schr\"odinger equation with a fractional Laplacian $(-\hbar^2 \Delta)^{\alpha/2}$ defined by the Riesz derivative \cite{lask00a,lask00b,lask00c,lask02}. The resulting dynamics are nonlocal in space and built on the L\'evy-path formulation, which is defined for $0 < \alpha \le 2$, leading to anomalous transport and modified energy spectra. Recent work has further clarified the mathematical properties and physical realizations of fractional Laplacians \cite{teym23}. It is worth mentioning that the Riesz derivative introduces spatial non-locality rather than an intrinsic temporal memory effect. The fractional kinetic operator couples distant spatial regions of the wavefunction through the non-local fractional Laplacian. In the path-integral formulation, standard quantum mechanics is associated with Brownian trajectories, while fractional quantum mechanics involves L\'evy trajectories characterized by heavy-tailed jump distributions. However, L\'evy flights are generally Markovian processes with independent increments and therefore do not imply memory effects. Consequently, the pseudo-eigenvalue equation arising from the Riesz derivative should be interpreted as a manifestation of spatial non-locality rather than as the origin of a temporal history-dependent dynamics.

Quantum mechanics in a fractional-dimensional space assumes that the effective dimension of space is noninteger ($D\notin \mathbb{N}$) \cite{sami12,rabe25}. This approach does not primarily modify derivatives but the geometric measure itself, allowing anisotropic, confined, or fractal-like media to be represented by an effective fractional dimension. Applications include semiconductor heterostructures, critical phenomena, and fractal systems.

The main difference is therefore conceptual: time-fractional evolution introduces temporal memory; space-fractional quantum mechanics introduces spatial nonlocality through a fractional kinetic operator; quantum mechanics in a fractional dimension modifies the geometry of the configuration space. Their connection lies in the common framework of fractional calculus and anomalous dynamics. All three theories seek to describe systems that deviate from ordinary quantum behavior due to memory effects, long-range correlations, or fractal geometry. Some studies even combine these different approaches, which may represent complementary aspects of quantum processes in complex media \cite{wang07,lask17,musl25}.

To our knowledge, there is currently no convincing experimental evidence that nature obeys fractional quantum mechanics (FQM) rather than ordinary quantum mechanics. On the other hand, there are experimental attempts to realize or simulate the dynamics predicted by FQM. A prominent example is an experimental realization of an optical fractional Schr\"odinger equation (FSE) for femtosecond laser pulses in the temporal domain \cite{shil23}. In \cite{ekra25}, solutions of FSE are computed for a possible future use to explore a wide range of important physical phenomena. The FSE is also an interesting theoretical tool for investigating various phenomena: tunneling and nonlocal transport phenomena in strong laser fields \cite{ciap26a,ciap26b}; role of disorder in the vibration spectra of molecules and atoms in solids \cite{step22}; equivalent models
of quantum gravity \cite{vara24}.  

Although not free from criticism \cite{yuch16,lask16}, the FSE introduced by Laskin \cite{lask00a,lask00b,lask00c,lask02,lask17} is the subject of this paper. In the event that nature had followed the path of FQM, it is expected that $\alpha$ must be very close to 2, otherwise deviations from usual quantum mechanics should already have been observed. It is also possible to consider a small deviation from the ordinary kinetic energy in phenomenological models. Our first purpose is to compute solutions of the FSE by considering $\alpha=2+ \epsilon$ with $|\epsilon| \ll 1$ (see Sect.~\ref{sec:Ham} for the justification for $\epsilon >0$). This study, which is new up to our knowledge, opens a way to possible comparisons with experiment. Note that perturbations of the FSE have been treated in \cite{bild18,bild20}, but the idea was to replace the orbital quantum number $l$ by $l+\alpha$ where $\alpha$ is presumed small, which is quite different from our approach. Our second objective is to compute the same solutions with the envelope theory (ET) \cite{hall83,sema13} which can produce approximate but reliable analytical eigensolutions for a wide variety of problems. The application of the ET for the FQM is also new. The use of both methods, perturbation theory and ET, enables a cross-check of the calculations and challenges the ET as a relevant method for treating fractional quantum equations. 

The framework of FQM considered and the definition of the perturbation calculations are given in Sect.~\ref{sec:Ham}. The ET is presented in Sect.~\ref{sec:ET}. The case of the harmonic oscillator is treated in Sect.~\ref{sec:HO}. This Hamiltonian is especially important because it can be used as a good approximation to study the lowest eigenvalues of most potential wells. The Kepler problem is treated in Sect.~\ref{sec:Kepler}. It is then possible to propose a connection with the experiment via the hydrogen atom.  Concluding remarks and outlook will be given in Sect.~\ref{sec:conclu}.

\section{Fractional quantum Hamiltonian}
\label{sec:Ham}

The time independent fractional Schr\"odinger Hamiltonian for a particle in the central 3-dimensional potential considered here is written 
\begin{equation}
\label{Halpha}
    H= D_\alpha p^\alpha + V(r),
\end{equation}
where ${\bm p}$ is the variable conjugate to the position ${\bm r}$, $p=|{\bm p}|$, $r=|{\bm r}|$, $D_\alpha$ is a constant, and the kinetic part is the fractional Laplacian \cite{kwas17,lisc20}
\begin{equation}
\label{Lap}
    p^\alpha = (\bm p^2)^{\alpha/2} = (-\hbar^2 \Delta)^{\alpha/2},
\end{equation}
defined by the Riesz derivative \cite{cai19} in the Laskin formulation. The Hamiltonian (\ref{Halpha}) is therefore nonlocal.

Several solutions of this Hamiltonian have already been found: free particle, Green function, Kepler problem, $1D$ power-law potential, infinite potential well, barrier penetration \cite{lask00c,lask02,guoa06,luch13,step19,sing23}. In this exploratory work, our purpose is to investigate small deviations from the ordinary Schr\"odinger Hamiltonian. To this end, we consider $\alpha=2+ \epsilon$ in (\ref{Halpha}) with $|\epsilon| \ll 1$, for two different central potentials. The generic one-body Hamiltonian considered has the following form
\begin{equation}
\label{HF}
    H_F= \frac{ p^{2+\epsilon} } {2 m \lambda^\epsilon} + V(r).
\end{equation}
where $\lambda$ is a positive constant with the dimension of a momentum. This constant is necessary to maintain the dimensional consistency of the Hamiltonian. Note that we can define a fundamental length $L_\lambda$ associated with the fundamental momentum $\lambda$ as
\begin{equation}
\label{Lalpha}
    L_\lambda= \frac{\hbar}{\lambda}.
\end{equation}

In the seminal work of Laskin \cite{lask00a}, FQM is built on the L\'evy-path formulation, which is defined for $0 < \alpha \le 2$. Nevertheless, $\alpha >2$ is not mathematically forbidden. In fact, some authors have explored generalized Hamiltonians for arbitrary positive values of $\alpha$ \cite{karp96,paus09,dona23}. These models can be physically meaningful as effective theories. However, they are usually viewed as higher-order dispersion models rather than as Laskin's FQM. In this paper, we consider both the negative (usual L\'evy-path formulation) and the positive (effective models) values for $\epsilon$. Since $|\epsilon|$ is very small, $H_F$ can be approximated by
\begin{equation}
\label{Hpert2}
    H_F\approx H_\epsilon = H_0 + \epsilon W(p),
\end{equation}
where
\begin{equation}
\label{Hpert3}
    H_0 = \frac{ {\bm p}^2 } {2 m} + V(r) \quad \textrm{and} \quad  W(p) = \frac{{\bm p}^2}{4 m}\ln \frac{{\bm p}^2}{\lambda^2}.
\end{equation}
Note that $W(p)\to 0$ when $p\to 0$. So $W(p)$ is regular at vanishing momentum. The problem can then be solved by the usual perturbation theory \cite{flug99,grif18,cohe19}, but with a perturbation that is momentum dependent. Actually, Hamiltonian~(\ref{Halpha}) can be solved indifferently in a configuration space or in a momentum space. For example, with $V(r)\propto r^2$, this problem treated with the momentum variable is mathematically equivalent to a Schr\"odinger equation with a power-law potential proportional to $r^\alpha$ treated with the position variable. This is particularly visible in a recent paper \cite{chev26}, in which the Lagrange-mesh method, initially developed in the configuration space, is adapted to direct calculations in momentum space. This is also clear in the formulation of the envelope theory (see Sect.~\ref{sec:ET}). The perturbation $W(p)$ grows faster than $p^2$. This is not a problem, since the case of a harmonic potential with cubic and quartic perturbations is mathematically analyzed in a correct manner in the configuration space \cite{flug99,grif18,cohe19}. Precautions must be taken for large values of quantum numbers. This will be discussed during the explicit calculations in Sect.~\ref{sec:HO} and \ref{sec:Kepler}.  
In the following, two different solvable Hamiltonians $H_0$ are considered and $\epsilon W(p)$ is treated as a perturbation. The computations will be limited to the determination of the first-order corrections in $\epsilon$ to the energies. 

\section{The envelope theory}
\label{sec:ET}

The ET is a method used to compute approximate solutions of $N$-body Hamiltonians in $D$ dimensions \cite{hall83,sema13}. The basic idea is to replace the initial Hamiltonian $H$ by an auxiliary Hamiltonian $\tilde H$, which is solvable. In practice, for $N\ge 3$ and $D\ge 2$, only harmonic oscillator Hamiltonians are usable \cite{will21}. For $N \le 2$ or $D=1$, other systems are possible \cite{flug99,calo71}. $\tilde H$ depends on parameters that are optimized in such a way that one of its particular eigenvalues is as close as possible to the corresponding eigenvalue of $H$. It is easy to implement for 1 particle, 2 different particles, or $N$ identical particles, but more complicated systems can also be treated. In some favorable cases, it is possible to determine approximate analytical eigenvalues, which can provide lower or upper bounds on exact eigenvalues \cite{cimi24}. In other systems, numerical approximation can always be computed. To lighten the paper, only the main properties of the ET are recalled. A description of the method can be found in \cite{sema18}, and a simpler pedagogical presentation is given in \cite{sema23}. 

In this paper, we only need to use the method for $N=1$ and $D=3$. Let us consider a Hamiltonian of this type
\begin{equation}
\label{HTV}
    H = T(p) + V(r).
\end{equation}
An approximate eigenvalue $E_{\textrm{ET}}$ is given by solving the following set of equations \cite{sema13}
\begin{align}
\label{ET1}
&E_{\textrm{ET}} = T(p_0) + V(r_0), \\
\label{ET2}
&r_0 p_0 = Q\hbar, \\
\label{ET3}
&p_0 T'(p_0) = r_0 V'(r_0),
\end{align}
where $U'(x) = dU(x)/dx$. If the auxiliary Hamiltoninan $\tilde H$ is a harmonic oscillator Hamiltonian, the global quantum number $Q$ is given by
\begin{equation}
\label{Q1}
Q = Q_\textrm{HO} = 2 n+l+ 3/2. 
\end{equation}
If $\tilde H$ is a Hamiltonian for the Kepler problem, then
\begin{equation}
\label{Q2}
Q = Q_\textrm{K} = n+l+ 1 .
\end{equation}
A state is specified by its radial ($n$) and orbital ($l$) quantum numbers. Once $r_0$ and $p_0$ are computed with (\ref{ET2})–(\ref{ET3}), $E_{\textrm{ET}}$ can be found with (\ref{ET1}). The parameter $r_0$ can be interpreted as the approximate mean distance between the particle and the origin, and $p_0$ as the approximate mean momentum of the particle. Note that (\ref{ET3}) is the translation of the master virial theorem \cite{luch90} for the ET. From the symmetry of (\ref{ET1})–(\ref{ET3}), it is clear that it is equivalent to work in position or in momentum variables. Approximations for the eigenvectors can also be computed with the ET, but this is not considered in this work.

The energies computed with the envelope theory are characterized by a strong unnatural degeneracy, as clearly shown by the structure of the global quantum numbers $Q_\textrm{HO}$ and  $Q_\textrm{K}$. This is inherent to the method because it relies on the solutions of an auxiliary Hamiltonian, which is solvable due to its high degree of symmetry. This drawback can be partly corrected by combining the ET with the dominantly orbital state method \cite{cimi24,sema15}. Significant improvements can then be obtained for some peculiar Hamiltonians, but the possible variational character of the ET solutions can be lost.  

Depending on the choice of $\tilde H$ (that is to say of $Q$) and the structure of $H$, the ET can provide a lower or upper bound of the exact eigenvalues in some favorable situations \cite{sema13}. In the following two sections, the only difference between the studied Hamiltonian $H$ and the chosen auxiliary Hamiltonian $\tilde H$ is the kinetic part. In this situation, the possibly variational character of the ET is only controlled by the function $b_T(x)$ such that \cite{sema13}
\begin{equation}
\label{bT}
   b_T(x) = T(\sqrt{x}) = \frac{x^{1+\epsilon/2}}{2 m \lambda^\epsilon},
\end{equation}
with $x \ge 0$. Since 
\begin{equation}
\label{bTpp}
   b_T(x)'' = \left( 1+ \frac{\epsilon}{2}\right) \frac{\epsilon}{2} \frac{x^{\epsilon/2-1}}{2 m \lambda^\epsilon}
\end{equation}
has always the same sign, an upper (lower) bound is obtained if $\epsilon < 0$ ($\epsilon > 0$).  

The ET is particularly suitable for perturbative calculations \cite{sema13}. The first step to solve the Hamiltonian $H_\epsilon$ is to find $E_{\textrm{ET}}$, $r_0$ and $p_0$ for the unperturbed Hamiltonian $H_0$ with the system (\ref{ET1})–(\ref{ET3}). The first order ET solution $E_\epsilon$ is then given in a second step by the computation of 
\begin{equation}
\label{ETfo}
   E_\epsilon = E_{\textrm{ET}} + \Delta E_{\textrm{ET}} \quad \textrm{with} \quad \Delta E_{\textrm{ET}}=\epsilon W(p_0).
\end{equation}
It is also possible to solve the system (\ref{ET1})–(\ref{ET3}) directly for $H_\epsilon$ and then perform a first-order expansion in $\epsilon$. However, this approach also leads to the result (\ref{ETfo}). As an inequality between two functions does not automatically hold for their finite-order limit expansions, it is worth noting that the existence of bounds for the full ET calculation does not guaranty their existence for the associated perturbative calculation. We will check below whether the variational character of the method is preserved for the results obtained. 

The effect of a perturbation is often to raise degeneracies. This is not possible with the ET, as previously reported. If the method has a variational character for the Hamiltonian studied, several non-degenerate levels that are characterized by the same global quantum number $Q$ will therefore have the same bound on the energy. It is perhaps possible to partly correct this problem by combining the ET with the dominantly orbital state method, but this is beyond the scope of this work.

Let us consider the following Hamiltonian, whose characteristics will be useful in the following sections.
\begin{equation}
\label{Halphabeta}
    H= D_\alpha p^\alpha + \textrm{sgn}(\beta) q^2 r^\beta,
\end{equation}
with $\alpha>0$ and $\beta > -\alpha$. Note that $\alpha=1$ with $D_\alpha=1$ corresponds to the kinetic energy of a massless particle, used in semirelativistic models for hadronic physics \cite{cimi24b}. The computation of $p_0$ gives
\begin{equation}
\label{p0alphabeta}
    p_0= \left( \frac{Q^\beta \hbar^\beta |\beta|q^2}{\alpha D_\alpha} \right)^{1/(\alpha+\beta)}.
\end{equation}
The ET eigenenergies are \cite{cimi24}
\begin{equation}
\label{Ealphabeta}
    E_{\textrm{ET}}= \textrm{sgn}(\beta)(\alpha+\beta)\left[ \left( \frac{D_\alpha}{|\beta|} \right)^\beta \left( \frac{q^2}{\alpha}\right)^\alpha \left(Q\hbar \right)^{\alpha\beta} \right]^{1/(\alpha+\beta)}.
\end{equation}
This formula is similar to the one obtained in \cite{lask02} by the WKB method for $D=1$. The ET is such that it gives the exact eigenenergies for a harmonic oscillator ($\alpha=\beta=2$) with $Q = Q_\textrm{HO}$, and for the Kepler problem ($\alpha=2,\beta=-1$) with $Q = Q_\textrm{K}$. When $0< \beta=\alpha < 2$ with $Q = Q_\textrm{HO}$, (\ref{Ealphabeta}) reduces to upper bounds for the eigenenergies of what is called the fractional generalization of the $3D$ harmonic oscillator Hamiltonian of standard quantum mechanics in \cite{lask00c}
\begin{equation}
\label{Ealphaalpha}
    E_{\textrm{ET}}= 2 \sqrt{D_\alpha q^2} \left((2 n+l+ 3/2)\hbar\right)^{\alpha/2} .
\end{equation}
The quality of the approximation can be improved by replacing the factor $2 n+l+ 3/2$ by the one predicted by the dominantly orbital state method, $\sqrt{2\alpha}\, n+l+ (1+\sqrt{2\alpha})/2$, but the variational character of the approximation is no longer guaranteed \cite{cimi24,sema15}. 

\section{Perturbative results for the harmonic oscillator}
\label{sec:HO}

Let us first look at the harmonic oscillator with
\begin{equation}
\label{VHO}
   V(r) = \frac{1}{2} m \omega^2 r^2.
\end{equation}
The eigenenergies are $E_{nl}=Q_\textrm{HO}\hbar \omega$ and the eigenfunctions $\psi_{nl}(\bm p)$ written in the momentum variable $\bm p$ are given in \cite{yane94}. The energy perturbations $\Delta E_{nl}$ are the expectation value $\langle \psi_{nl}(\bm p)|\epsilon  W(p)|\psi_{nl}(\bm p)\rangle$. These integrals can be computed exactly using the Feynman trick \cite{nahi15}. An example of computations is given in the Appendix~\ref{sec:app}. The closed-form expressions derived in this section were checked by direct numerical integration of the expectation values for various parameter values. No general formula has been obtained, but the same structure appears for all tested $\{n,l\}$ sets: 
\begin{equation}
\label{PHO}
   \Delta E_{nl} = \epsilon\frac{Q_\textrm{HO}}{4} \ln \left( \frac{e^{A_{nl}-\gamma}}{4} \frac{m\hbar\omega}{\lambda^2} \right)\hbar \omega,
\end{equation}
where $\gamma$ is the Euler constant and where $A_{nl}$ is a coefficient depending on $\{n,l\}$, some of whose values are given in Table~\ref{tab:Anl}. As expected, the replacement of $\bm p^2$ by $p^{2+\epsilon}$ has removed the strong degeneracy of the harmonic oscillator. Two different $\{n,l\}$ sets corresponding to the same value of $Q_\textrm{HO}$ are given to show the removal of degeneracy. 

\begin{table}[!ht]
\centering 
\begin{tabular}{cccc} 
\hline\hline
 ${n\ l}$ & $Q_\textrm{HO}$ & $A_{nl}$ & $e^{A_{nl}-\gamma}/4$ \\ 
\hline
0\ 0 & 3/2 & 8/3 & 2.020 \\ 
1\ 0 & 7/2 & 74/21 & 4.760 \\
0\ 2 & 7/2 & 352/105 & 4.010 \\ 
5\ 0 & 23/2 & 375598/79695 & 15.63 \\
0\ 10 & 23/2 & 1488711776/334639305 & 12.00 \\ 
\hline\hline
\end{tabular} 
\caption{$A_{nl}$ values for some $\{n,l\}$ sets with $Q_\textrm{HO}$ and numerical values of the characteristic coefficient in the ln-term of (\ref{PHO}).}
\label{tab:Anl}
\end{table}

The eigensolutions of the Hamiltonian~(\ref{HF}) with potential~(\ref{VHO}) can be computed with a high accuracy thanks to the Lagrange mesh method \cite{baye15}: relative errors below $10^{-8}$ can easily be obtained. So, it is possible to check numerically the precision of formula~(\ref{PHO}). In the case $\hbar=m=\omega=1$, the relative error for the ground state energy between the exact solution and the perturbation approximation reach a maximum of 17\% for $|\epsilon| = 0.1$ when $\lambda \in [10^{-4}, 10^4]$. This maximal error drops below 0.001\% when $|\epsilon| = 0.001$. When $\lambda \in [0.1, 10]$ with $|\epsilon| = 0.1$, the relative errors reach a maximum of 0.8\% for the ground state and 1.5\% for states with $Q_\textrm{HO}=23/2$. Hence, formula~(\ref{PHO}) seems relevant for a wide range of $\epsilon$ and $m\hbar\omega/\lambda^2$ values. However, the quality of the perturbation approximation decreases as the excitation energy increases. A similar behavior occurs for quartic perturbations of the harmonic oscillator \cite{grif18,cohe19}.

The first step of the ET calculation is to compute $p_0$, with (\ref{ET2})–(\ref{ET3}), for the Hamiltonian $H_0$ with (\ref{VHO}). From (\ref{p0alphabeta}) and (\ref{Ealphabeta}), we have
\begin{equation}
\label{p0HO}
   p_0 = \sqrt{m Q \hbar \omega}.
\end{equation}
With $Q = Q_\textrm{HO}$, the ET solution (\ref{ET1}) is the exact value $E_{nl}^{\textrm{ET}}=E_{nl}=Q_\textrm{HO}\hbar\omega$. Using (\ref{ETfo}), the contribution of the perturbation is 
\begin{equation}
\label{PHOET}
   \Delta E_{nl}^{\textrm{ET}} = \epsilon\frac{Q_\textrm{HO}}{4}  \ln \left( Q_\textrm{HO}\frac{m\hbar\omega}{\lambda^2} \right)\hbar\omega,
\end{equation}
Formulas~(\ref{PHO}) and (\ref{p0HO}) have a similar structure. The only difference is the replacement of $e^{A_{nl}-\gamma}/4$ by $Q_\textrm{HO}$ in the ln-term. Comparisons between these two last quantities for some sets $\{n,l\}$ are given in Table~\ref{tab:HO_ET} with the ratio $R_{nl}=(e^{A_{nl}-\gamma}/4)/Q_\textrm{HO}$. All the values of $R_{nl}$ are close to 1, indicating that the ET approximation is quite good for the lowest energy states. Since the ln-function is a monotonically increasing function, $\Delta E_{nl}^{\textrm{ET}}$ is an upper (lower) bound of $\Delta E_{nl}$ if $\epsilon <0$ ($\epsilon >0$) for all the cases computed. So, the variational character expected from (\ref{bTpp}) is preserved. Note that the absolute error $| \Delta E_{nl}- \Delta E_{nl}^{\textrm{ET}}|$ does not depend on $\lambda$ due to the ln-term, but this is not the case for the relative error. 

\begin{table}[!ht]
\centering 
\begin{tabular}{cccc} 
\hline\hline
 ${n\ l}$ & $Q_\textrm{HO}$ & $e^{A_{nl}-\gamma}/4$ & $R_{nl}$\\
\hline
0\ 0 & 1.5 & 2.020 & 1.35 \\
1\ 0 & 3.5 & 4.760 & 1.36 \\ 
0\ 2 & 3.5 & 4.010 & 1.15 \\ 
5\ 0 & 11.5 & 15.63 & 1.36 \\ 
0\ 10 & 11.5 & 12.00 & 1.04 \\ 
\hline\hline
\end{tabular} 
\caption{Comparisons of formulas~(\ref{PHO}) and (\ref{p0HO}) for some sets $\{n,l\}$. The last column is the ratio $R_{nl}=(e^{A_{nl}-\gamma}/4)/Q_\textrm{HO}$.} 
\label{tab:HO_ET}
\end{table}

Some central potentials $U(r)$ can be developed around the radial position $r_m$ of their minimum, 
\begin{equation}
\label{hoapprx}
   U(r) = U(r_m) + \frac{1}{2} U''(r_m) (r-r_m)^2.
\end{equation} 
For lower-energy states, the wave function is localized near the minimum, and higher-order terms of the potential $U(r)$ are often negligible. So, good approximations of lowest vibrational energies can be obtained as the eigenvalues of the harmonic potential~(\ref{hoapprx}) \cite{grif18,cohe19}. This is the case for the Morse potential and the Lennard-Jones potential, which are often used to compute the spectra of a diatomic molecule. If $r_m=0$, as in the case of Gaussian potentials used to model atomic traps, all the lowest states, vibrational and orbital, can be approximated by the eigenvalues of~(\ref{hoapprx}). In these situations, an approximation of the perturbation (\ref{PHOET}) for a potential well can be obtained by replacing $\omega$ by $\sqrt{ U''(r_m)/m}$.

\section{Perturbative results for the Kepler problem}
\label{sec:Kepler} 

The Kepler problem corresponding to the $2D$ hydrogen atom is treated in \cite{step19}. The WKB solutions of the same $3D$ system are computed in \cite{lask00c,lask02}. In this section, we repeat the procedure of the previous one for the hydrogen atom with
\begin{equation}
\label{VHA}
   V(r) = -\frac{1}{r},
\end{equation}
written in atomic units. Within the same units, the eigenenergies\footnote[1]{To avoid overloading the text with symbols, the notations used in the previous section ($E_{nl}$, $p_0$, \ldots) will also be used unchanged in this one.} are $E_{nl}=-1/(2 Q_\textrm{K}^2)$ and the eigenfunctions $\psi_{nl}(\bm p)$ written in the momentum variable $\bm p$ are given in \cite{yane94}. The energy perturbations $\Delta E_{nl}$ are the expectation value $\langle \psi_{nl}(\bm p)|\epsilon  W(p)|\psi_{nl}(\bm p)\rangle$. Again, these integrals can be computed exactly using the Feynman trick \cite{nahi15}. An example of computations is given in the Appendix~\ref{sec:app}. Calculations were also checked by direct numerical integration of the expectation values for various $\lambda$ values. No general formula has been obtained, but the same structure appears for all tested $\{n,l\}$ sets: 
\begin{equation}
\label{PHA}
   \Delta E_{nl} = \epsilon\frac{1}{2 Q_\textrm{K}^2} \ln \left( \frac{e^{B_{nl}}}{Q_\textrm{K} \lambda} \right),
\end{equation}
where $B_{nl}$ is a coefficient depending on $\{n,l\}$, some of whose values are given in Table~\ref{tab:Bnl}. As expected, the replacement of $\bm p^2$ by $p^{2+\epsilon}$ has removed the strong degeneracy of the Kepler problem. Two different $\{n,l\}$ sets corresponding to the same value of $Q_\textrm{K}$ are given to show the removal of degeneracy, which is relatively larger than for the harmonic oscillator case. Let us recall that $n$ is the radial quantum number, the principal quantum number being $Q_\textrm{K}$. 

\begin{table}[!ht]
\centering 
\begin{tabular}{cccc} 
\hline\hline
 ${n\ l}$ & $Q_\textrm{K}$ & $B_{nl}$ & $e^{B_{nl}}/Q_\textrm{K}$ \\ 
\hline
0\ 0 & 1 & 1/3 & 1.396 \\ 
1\ 0 & 2 & 11/15 & 1.041 \\
0\ 1 & 2 & 1/5 & 0.6107 \\ 
5\ 0 & 6 & 131/143 & 0.4166 \\ 
0\ 5 & 6 & 1/13 & 0.1800 \\ 
\hline\hline
\end{tabular} 
\caption{$B_{nl}$ values for some $\{n,l\}$ sets with $Q_\textrm{K}$ and numerical values of the characteristic coefficient in the ln-term of (\ref{PHA}).}
\label{tab:Bnl}
\end{table}

As in the case of the harmonic potential, the eigensolutions of the Hamiltonian~(\ref{HF}) with the potential~(\ref{VHA}) can be calculated with high precision thanks to the Lagrange mesh method, and the precision of formula~(\ref{PHA}) can be checked numerically. When $\lambda \in [0.1, 10]$ with $|\epsilon| = 0.1$, the relative errors stay below 8\% for states with $Q_\textrm{K}\le 6$. This error is divided by more than 100 for $|\epsilon| = 0.001$ within the same range of $\lambda$. So, formula~(\ref{PHA}) seems relevant for a significative range of $\epsilon$ and $\lambda$ values. As in the case of the harmonic oscillator, the quality of the perturbation approximation decreases as the excitation energy increases. 

Evaluating $p_0$ for the Hamiltonian $H_0$ in atomic units with (\ref{VHA}) using (\ref{ET2})–(\ref{ET3}), we have, from (\ref{p0alphabeta}) and (\ref{Ealphabeta}),
\begin{equation}
\label{p0HA}
   p_0 = \frac{1}{Q}.
\end{equation}
If $Q=Q_\textrm{K}$, the ET solution (\ref{ET1}) is the exact value $E_{nl}^{\textrm{ET}}=E_{nl}=-1/(2Q_\textrm{K}^2)$. Using (\ref{ETfo}), the contribution of the perturbation is 
\begin{equation}
\label{PHAET}
   \Delta E_{nl}^{\textrm{ET}} = \epsilon\frac{1}{2Q_\textrm{K}^2}  \ln \left( \frac{1}{Q_\textrm{K} \lambda} \right).
\end{equation}
Formulas~(\ref{PHA}) and (\ref{p0HA}) have a similar structure. The only difference is the replacement of $e^{B_{nl}} Q_\textrm{K}^{-1}$ by $Q_\textrm{K}^{-1}$ in the ln-term. Comparisons between these two last quantities for some $\{n,l\}$ sets are given in Table~\ref{tab:HA_ET} with the values of the ratio $R_{nl}= e^{B_{nl}} Q_\textrm{K}^{-1}/Q_\textrm{K}^{-1}=e^{B_{nl}}$. These values are reasonable but sensibly greater that in the case of the harmonic oscillator system. This is not really surprising, since a correlation between the presence of a divergence and a decrease in the accuracy of the ET was demonstrated in \cite{cim24b}, albeit in a different context. As in the previous section, $\Delta E_{nl}^{\textrm{ET}}$ is an upper (lower) bound of $\Delta E_{nl}$ if $\epsilon <0$ ($\epsilon >0$) for all the cases computed. So, the variational character expected from (\ref{bTpp}) is preserved. Again, the absolute error $| \Delta E_{nl}- \Delta E_{nl}^{\textrm{ET}}|$ does not depend on $\lambda$ due to the ln-term, but this is not the case for the relative error. 

\begin{table}[!ht]
\centering 
\begin{tabular}{cccc} 
\hline\hline
 ${n\ l}$ & $Q_\textrm{K}$ & $e^{B_{nl}}/Q_\textrm{K}$ & $R_{nl}$\\
\hline
0\ 0 & 1 & 1.396 & 1.4 \\ 
1\ 0 & 2 & 1.041 & 2.1 \\ 
0\ 1 & 2 & 0.6107 & 1.2 \\ 
5\ 0 & 6 & 0.4166 & 2.5 \\
0\ 5 & 6 & 0.1800 & 1.1 \\ 
\hline\hline
\end{tabular} 
\caption{Comparisons of formulas~(\ref{PHA}) and (\ref{p0HA}) for some sets $\{n,l\}$. The last column is the ratio $R_{nl}=e^{B_{nl}}$.} 
\label{tab:HA_ET}
\end{table}

As the above results are related to the hydrogen atom, it seems feasible to obtain information about the possible values of the parameters $\epsilon$ and $\lambda$, in the event that nature deviates slightly from what is generally expected. Equation~(\ref{PHA}) can be rewritten
\begin{equation}
\label{PHA2}
   \frac{\Delta E_{nl}}{E_{nl}} = -\epsilon \ln \left( \frac{e^{B_{nl}}}{Q_\textrm{K} \lambda} \right).
\end{equation}
We can expect this dimensionless quantity to be much less than 1, otherwise deviations from the usual quantum mechanics should already have been observed. To obtain reliable phenomenological estimations, it is necessary to compute the effect of the fractional corrections on several observables, such as accurate transition energies, and compare them with fine experimental measurements, such as QED corrections. This huge work is left for further studies. Nevertheless, we can try to propose a very crude bound on the parameters $\epsilon$ and $\lambda$. A recent measurement of the ground state energy of the hydrogen atom shows that the relative uncertainty about this quantity is around $2\times 10^{-12}$ \cite{sche24}. This suggests that 
\begin{equation}
\label{PHA200}
  \left| \frac{\Delta E_{00}}{E_{00}} \right|= \left|\epsilon \ln \left( e^{1/3}L_\lambda \right)\right| \lesssim 2\times 10^{-12},
\end{equation}
with $L_\lambda=1/\lambda$ in units of Bohr radius. For example, $\ln \left( e^{1/3}L_\lambda \right)$ varies between $-8.88$ and $9.54$ when $L_\lambda$ varies from $10^{-4}$ to $10^4$.

\section{Concluding remarks}
\label{sec:conclu}

The fractional Schrödinger equation has been investigated as a small perturbation of the standard Schrödinger equation by introducing a kinetic-energy operator that deviates slightly from the usual nonrelativistic form. To ensure dimensional consistency, a positive constant with the dimensions of momentum (or inverse length) has been introduced. In this exploratory study, two illustrative systems, the harmonic oscillator and the Kepler problem, have been examined. For the latter, a potential connection with experimental observations has been suggested.

For both Hamiltonians, analytical expressions for the lowest-energy states have been derived within the framework of perturbation theory. Their accuracy has been validated through numerical calculations for various parameter values. Analytical approximations for the complete spectra have also been obtained using envelope theory. For all states investigated, the envelope theory eigenenergies provide either upper or lower bounds. These bounds are generally of good quality for the harmonic oscillator across the spectrum. In the Kepler case, the agreement remains satisfactory, but less accurate than for the harmonic oscillator.

A natural many-body extension of this work is to consider the following Hamiltonian,
\begin{equation}
\label{HNB}
    H = \sum_{i=1}^N \frac{|\bm p_i|^{2+\epsilon}}{2 m} + \sum_{i<j=2}^N V(|\bm r_i - \bm r_j|) \quad \textrm{with} \quad \sum_{i=1}^N \bm p_i = \bm 0,
\end{equation}
whose exact solutions are known for $\epsilon=0$, $V(z) \propto z^2$ and arbitrary values of $N$. Envelope theory can readily be extended to many-body systems with nonstandard kinetic-energy operators, particularly when all the particles are identical. It therefore offers a practical framework for obtaining approximate eigensolutions of many-body Hamiltonians arising in fractional quantum mechanics, like~(\ref{HNB}). A reasonable level of accuracy may be expected for harmonic-oscillator systems. Although its performance for more general interactions remains to be assessed, the method deserves further investigation given the intrinsic difficulty of solving many-body quantum systems.

The spectra obtained with envelope theory exhibit strong spurious degeneracy. This feature is intrinsic to the method, which relies on an auxiliary Hamiltonian whose solvability stems from its high degree of symmetry. This drawback can be partially corrected by combining the envelope theory with the dominant orbital state method, leading in some cases to significant improvements. Exploring such hybrid approaches in the context of fractional quantum mechanics appears to be a promising direction for future work.

Since conventional perturbation theory yields accurate predictions for the eigenenergies of the fractional Schrödinger equation, it would be worthwhile to extend the analysis to wavefunctions and derived observables. In the case of the hydrogen atom, such calculations could provide a more direct assessment of the physical relevance of fractional quantum mechanics and, ultimately, offer a means of confronting the theory with high-precision experimental data.

\section*{Acknowledgements}
C.T. has received support from the Communauté française de Belgique as part of the funding for a FRIA grant. This work was also supported by the IISN under Grant Number 4.45.10.08. We would also like to thank anonymous referees who suggested numerous improvements, as well as S\'ebastien Fumeron for his advices.

\appendix

\section{Examples of quadrature}
\label{sec:app}

The ground state of the harmonic oscillator in the momentum variable is given by \cite{yane94}
\begin{equation}
\label{A0}
    \psi_{00}(\bm p) = \frac{1}{(m\pi\hbar\omega)^{3/4}} \exp\left( -\frac{p^2}{2 m\hbar\omega} \right).
\end{equation}
To evaluate $\langle \psi_{00}(\bm p)|W(p)|\psi_{00}(\bm p)\rangle$, by omitting a constant factor, one must compute
\begin{align}
\label{A1}
    \int_0^\infty p^4 \exp{\left( -\frac{p^2}{m\hbar\omega} \right)} \ln \frac{p^2}{\lambda^2} dp &= \frac{2}{\delta^{5/2}} \int_0^\infty z^4 e^{-z^2} \ln \frac{z}{\sqrt{\delta}\lambda} dz \nonumber\\
    &= \frac{2}{\delta^{5/2}} \left( \int_0^\infty z^4 e^{-z^2}  \ln z\, dz - \ln(\sqrt{\delta}\lambda) \int_0^\infty z^4 e^{-z^2} dz \right), 
\end{align}
with $\delta=(m\hbar\omega)^{-1}$. The last integral in (\ref{A1}) is a standard one and equals $3\sqrt{\pi}/8$. The penultimate one is more tricky but can be computed with the Feynman trick. One can write
\begin{align}
\label{A2}
    4 \int_0^\infty z^{4\theta} e^{-z^2}  \ln z\, dz  &=
    \int_0^\infty t^{2\theta-1/2} e^{-t}  \ln t\, dt \nonumber \\
    & = \frac{1}{2}\frac{\partial}{\partial \theta} \Gamma(2\theta +1/2) \nonumber \\
    & = \Gamma(2\theta +1/2)\, \psi(2\theta +1/2),
\end{align}
where $\psi(z)$ is the digamma function. So,
\begin{equation}
\label{A3}
    \int_0^\infty z^4 e^{-z^2}  \ln z\, dz = \frac{1}{4}\Gamma(5/2)\, \psi(5/2)=\frac{\sqrt{\pi}}{16} \left(  8-6 \ln 2-3 \gamma\right),
\end{equation}
where $\gamma$ is the Euler's constant.

A similar procedure is used to determine $\langle \psi_{00}(\bm p)|W(p)|\psi_{00}(\bm p)\rangle$ for the hydrogen atom. In atomic units, its ground state in the momentum variable is given by \cite{yane94}
\begin{equation}
\label{A0b}
    \psi_{00}(\bm p) = \frac{2\sqrt{2}}{\pi} \frac{1}{\left( 1+p^2 \right)^2}.
\end{equation}
By omitting a constant factor, it is then necessary to compute
\begin{equation}
\label{A4}
    \int_0^\infty p^4 \frac{1}{\left( 1+p^2 \right)^4} \ln \frac{p}{\lambda} dp =   \int_0^\infty  \frac{p^4 \ln p}{\left( 1+p^2 \right)^4} dp
    - \ln \lambda \int_0^\infty  \frac{p^4}{\left( 1+p^2 \right)^4} dp.
\end{equation}
The last integral in (\ref{A4}) is a standard one and equals $\pi/32$, while the penultimate one can be computed again with the Feynman trick. One can write 
\begin{align}
\label{A5}
   4 \int_0^\infty  \frac{p^{4\theta} \ln p}{\left( 1+p^2 \right)^4} dp & = 
   \frac{\partial}{\partial \theta} \int_0^\infty  p^{4\theta}\left( 1+p^2 \right)^{-4} dp \nonumber \\ 
   & =\frac{1}{2} \frac{\partial}{\partial \theta} \int_0^\infty  z^{2\theta-1/2}\left( 1+z \right)^{-4} dz\nonumber \\
   & =\frac{1}{2} \frac{\partial}{\partial \theta} B\left( 2\theta-1/2,7/2-2\theta \right),
\end{align}
Where $B(x,y)$ is the beta function. The evaluation of these formulas for $\theta=1$ finally gives
\begin{equation}
\label{A6}
    \int_0^\infty  \frac{p^4 \ln p}{\left( 1+p^2 \right)^4} dp = \frac{\pi}{96}.
\end{equation}

\end{document}